\documentclass[conference]{IEEEtran}
\ifCLASSINFOpdf
   \usepackage[pdftex]{graphicx}
   \DeclareGraphicsExtensions{.pdf,.jpeg,.png}
\else
\fi
\usepackage{amsmath,dsfont,bbm,epsfig,amssymb,amsfonts,amstext,verbatim}\usepackage{amsopn,cite,subfigure}
\usepackage{multirow,multicol,lipsum,xfrac}
\usepackage{amsthm}
\usepackage{mathtools,amsthm}
\usepackage{perpage}
\usepackage{balance}
\usepackage{url}
\usepackage{amsfonts}
\usepackage{epsfig}
\usepackage[font={small}]{caption}
\usepackage{graphicx}
\usepackage{psfrag}	
\usepackage{etoolbox}
\usepackage{algorithmicx}
\usepackage[Algorithm,ruled]{algorithm}
\usepackage{algpseudocode}
\usepackage{pifont}
\usepackage[utf8]{inputenc}
\usepackage[T1]{fontenc}  
\usepackage[nolist]{acronym}
\usepackage{paralist}
\usepackage{enumitem}
\usepackage{bbm}
\usepackage[process=auto]{pstool}
\usepackage{tikz,pgfplots}
\usetikzlibrary{shapes,arrows}
\usepackage{pdfpages}

\newcommand{\setR}{\mathbb{R}}

\newcommand{\setC}{\mathbb{C}}

\newcommand{\setL}{\mathbb{L}}

\newcommand{\snr}{\mathrm{SNR}}
\newcommand{\distB}[2]{\mathcal{B} \left( #1,#2 \right) }

\newcommand{\rms}{\mathrm{s}}

\newcommand{\sfC}{\mathsf{C}}

\newcommand{\her}{\mathsf{H}}

\newcommand{\mar}{\mathcal{R}}

\newcommand{\man}{\mathcal{N}}

\newcommand{\mm}{\mathrm{m}}
\newcommand{\ee}{\mathrm{e}}

\newcommand{\bh}{{\mathbf{h}}}
\newcommand{\tbh}{{\mathbf{\tilde{h}}}}

\newcommand{\bx}{{\boldsymbol{x}}}

\newcommand{\set}[1]{\left\lbrace#1\right\rbrace}

\newcommand{\bg}{{\mathbf{g}}}
\newcommand{\bs}{{\boldsymbol{s}}}

\newcommand{\bv}{{\boldsymbol{v}}}

\newcommand{\bww}{\mathbf{w}}

\newcommand{\tbg}{\mathbf{\tilde{g}}}

\newcommand{\by}{{\boldsymbol{y}}}

\newcommand{\bn}{{\boldsymbol{n}}}

\newcommand{\trp}{\mathsf{T}}

\newcommand{\mW}{\mathbf{W}}
\newcommand{\tmH}{\mathbf{\tilde{H}}}

\newcommand{\mI}{\mathbf{I}}

\newcommand{\mG}{\mathbf{G}}

\newcommand{\mGam}{\mathbf{\Gamma}}
\newcommand{\mU}{\mathbf{U}}

\newcommand{\mV}{\mathbf{V}}

\newcommand{\mH}{\mathbf{H}}

\newcommand{\E}{\mathbb{E}\hspace{.5mm}}

\newcommand{\sinr}{{\mathrm{SINR}}}

\newcommand{\norm}[1]{\lVert #1 \rVert}

\newcommand{\abs}[1]{\lvert #1 \rvert}
\newcommand{\tr}[1]{\mathrm{Tr} \{ #1 \}}

\newtheoremstyle{mystyle}%                % Name
  {}%                                     % Space above
  {}%                                     % Space below
  {}%                                     % Body font
  {}%                                     % Indent amount
  {\bfseries}%                            % Theorem head font
  {:}%                                     % Punctuation after theorem head
  { }%                                    % Space after theorem head, ' ', or \newline
  {}%                                     % Theorem head spec (can be left empty, meaning `normal')

\theoremstyle{mystyle}

\newtheorem{definition}{Definition}

\newtheorem{remark}{Remark}

\algnewcommand\algorithmicLet{\textbf{Let}}
\algnewcommand\Let{\item[\algorithmicLet]}
\algnewcommand\algorithmicSet{\textbf{Set}}
\algnewcommand\Set{\item[\algorithmicSet]}

\algnewcommand\algorithmicInitiate{\textbf{Initiate}}
\algnewcommand\Initiate{\item[\algorithmicInitiate]}
\algnewcommand\algorithmicStart{\textbf{Begin}}
\algnewcommand\Begin{\item[\algorithmicStart]}
\algnewcommand\algorithmicEnd{\textbf{End}}
\algnewcommand\End{\item[\algorithmicEnd]}

\algnewcommand\algorithmicOutP{\textbf{Output:}}
\algnewcommand\Out{\item[\algorithmicOutP]}

\algnewcommand\algorithmicInP{\textbf{Input:}}
\algnewcommand\In{\item[\algorithmicInP]}

\newcounter{bar}

\begin{document}
\title{On Robustness of Massive MIMO Systems Against Passive Eavesdropping under Antenna~Selection}

% Authors
\author{
\IEEEauthorblockN{
Ali Bereyhi\IEEEauthorrefmark{1},
Saba Asaad\IEEEauthorrefmark{1}\IEEEauthorrefmark{2},
Ralf R. M\"uller\IEEEauthorrefmark{1},
Rafael F. Schaefer\IEEEauthorrefmark{3},
Amir M. Rabiei\IEEEauthorrefmark{2}
}
\IEEEauthorblockA{
\IEEEauthorrefmark{1}Institute for Digital Communications (IDC), Friedrich-Alexander Universit\"at Erlangen-N\"urnberg (FAU)\\
\IEEEauthorrefmark{3}Information Theory and Applications Chair, Technische Universit\"at Berlin (TUB)\\
\IEEEauthorrefmark{2}School of Electrical and Computer Engineering, University of Tehran\\
ali.bereyhi@fau.de, saba{\_}asaad@ut.ac.ir, ralf.r.mueller@fau.de, rafael.schaefer@tu-berlin.de, rabiei@ut.ac.ir
\thanks{This work has been accepted for presentation in IEEE Global Communications Conference (Globecom) 2018 in Abu Dhabi, UAE. The link to the final version in the Proceedings of Globecom will be available later.}
\thanks{This work was supported by the German Research Foundation, Deutsche Forschungsgemeinschaft (DFG), under Grant No. MU 3735/2-1.}
}
}
%\IEEEspecialpapernotice{(Invited Paper)}
\IEEEoverridecommandlockouts

% make the title area
\maketitle

\begin{abstract}
In massive~MIMO wiretap settings, the base station can significantly suppress eavesdroppers by narrow beamforming toward legitimate terminals. Numerical investigations show that by this approach, secrecy is obtained at no significant cost. We call this property of massive MIMO systems ``secrecy for free'' and show that it not only holds when all the transmit antennas at the base station are employed, but also when only a \textit{single} antenna is set active. Using linear precoding, the information leakage to the eavesdroppers can be sufficiently diminished, when the \textit{total} number of available transmit antennas at the base station grows large, even when only a \textit{fixed} number of them are selected. This result indicates that passive eavesdropping has no significant impact on massive MIMO systems, regardless of the number of active transmit antennas.\vspace*{1mm}
%
%Our investigations demonstrate that even with a fixed number of active transmit antennas at the base station, the information leakage to the eavesdroppers can be suppressed in the large-system limit by simple linear precoding. The achievable rate over the legitimate channel in this case grows proportional to the \textit{total} number of transmit antennas. 
%
%This result indicates that passive eavesdropping has no significant impact in massive MIMO systems, and thus, the secrecy rate is close to the rate achieved over the legitimate channel.
%
%
%
% considering both scenarios of selecting the total available antennas and a finite number of them. In particular, we show that both scenarios are robust from passive eavesdropping. In particular, passive eavesdropping has no effect on the beamforming due to the asymptotic orthogonality of wireless channel in large dimension. Although the analysis are given in large system limit, the numerical result indicates the information leakage converges to zero, even in not so large dimension. Note that under antenna selection, secrecy for free take place at a slower rate. 
%
 
\end{abstract}

\IEEEpeerreviewmaketitle

\section{Introduction}
Recently, massive \ac{mimo} systems have emerged as a promising enabling technology to address the explosive growth of data traffic in the next generation of wireless networks (5G) \cite{boccardi2014five}. Reliable and secure transmission of data in 5G is of paramount importance for system designers. In this respect, physical layer~security,~complemented with cryptographic approaches in upper layers of the network, provides a well-integrated secure platform by exploiting the imperfection of the communication channels \cite{yang2015safeguarding}. The pioneering work on physical layer security goes back to Wyner who studied a point to point wiretap channel in \cite{wyner1975wire} and showed that 
confidential message transmission is possible as long as the eavesdropper observes a degraded version of the signal received at the legitimate terminal. Wyner's result was later extended to several other settings including \ac{mimo} wiretap channels \cite{khisti2010secure, oggier2011secrecy}.

In massive \ac{mimo} systems, the large number of antennas can provide more security by narrow beamforming toward legitimate receivers. In this case, the signal strength at the eavesdropper, which is located somewhere outside the main beam, is much lower than the strength of the signal received at the legitimate terminal. This observation was demonstrated in \cite{kapetanovic2015physical} via numerical investigations where the authors showed that passive eavesdropping in not-too-dense networks has little impacts on the secrecy performance. Similar to other performance gains of massive \ac{mimo}, this robustness against passive eavesdropping is obtained at the expense of~high~hardware cost and complexity imposed by the large number of antennas in these systems. The impact of passive eavesdropping on~the secrecy performance in dense networks has been~further~investigated in \cite{wang2016secrecy}. %In contrast to the case with passive eavesdroppers, the achievable secrecy rate was shown in \cite{kapetanovic2015physical} to dramatically reduce when the eavesdropper actively overhears the channel between the \ac{bs} and the legitimate terminals. %In \cite{zhu2014secure} a secure transmission scheme is investigated for both perfect and imperfect \ac{csi} by generating artificial noise in a muti-cell multi-user system. 
%Employing a large number of antennas incurs high hardware cost and complexity due to the 
%In fact, in these systems each transmit antenna needs a dedicated \ac{rf}-chain. Consequently, by employing a large array of antennas for transmission the overall \ac{rf}-cost 
%As the result, 

Several approaches were proposed in the literature to alleviate the cost-complexity issue of massive \ac{mimo}, e.g., \cite{mohammed2013per,asaad2017tas,bereyhi2017asymptotics}. A promising solution is antenna selection in which the transmission is carried out through a subset of antennas; see \cite{benmimoune2015joint} and \cite{bereyhi2018stepwise}, and the references therein. In addition to its main objective, i.e., reducing the overall \ac{rf}-cost, antenna selection has shown to enhance the performance in some scenarios. In \cite{bereyhi2018stepwise} and \cite{li2014energy}, it was demonstrated that in several \ac{mimo} settings, energy efficiency is not an increasing function of the number of transmit antennas, and thus, it can be improved by switching off some of them.
The authors in \cite{asaad2017optimal}, moreover, showed that in \ac{mimo} wiretap settings, the secrecy performance, when no precoding is utilized at the \ac{bs}, is optimized when only a few transmit antennas are set active. The security benefits of optimal single antenna selection was further discussed in \cite{zhu2016secrecy}.
\subsection*{Objectives and Contributions}
Considering the implementational issues in massive \ac{mimo} systems, this study aims to answer the following question:~Are massive \ac{mimo} wiretap settings robust against passive eavesdropping when only a subset of transmit antennas are set active? To this end, we investigate the robustness against passive eavesdropping by defining the concept of ``secrecy for free''. It is then shown that even when only a~\textit{single}~transmit antenna is set active, by simple linear precoding, the information leakage to the eavesdropper vanishes, as the \textit{total} number of available antennas at the \ac{bs} increases. This result indicates that with a large transmit antenna array, regardless of the number of active antennas, secrecy can be achieved at no significant~cost. Our analysis moreover provides rigorous justifications for the earlier numerical observations on the robustness of massive \ac{mimo} systems against passive eavesdropping, e.g., \cite{kapetanovic2015physical}, considering a more generic setup. %with antenna selection.

%We call this generic property of massive \ac{mimo} systems ``secrecy for free'' under passive eavesdropping and  show that it holds in the asymptotic regime even when a \textit{fixed} number of antennas are selected. Our analyses are further confirmed via numerical investigations.
% and antenna selection one. We consider a general multi-user \ac{mimo} wiretap channel with a large number of antennas at the transmitter. We have shown that in the case of full complexity, i.e., the case in which the number of selected antennas equals to the total number of selected antennas, due to the favorable propagation property of massive \ac{mimo} systems, the information leakage to eavesdropper converges to zero. . Here, we consider a \ac{tas} protocol which selects antennas correspond to the strongest channel gains. Our results show that in the asymptotic regime with finite number of antennas, the achievable rate in the absence of the eavesdropper is nearly equal to the secrecy rate in the presence of an eavesdropper, i.e., passive eavesdropper has a negligible effect on secrecy performance even with a fixed number of active antennas.    
% The investigations consider a general framework which encloses both the scenarios of full complexity, i.e., the case in which all transmit antennas are set active, and antenna selection.  
\subsection*{Notations}
Throughout the paper, scalars, vectors and matrices~are~represented by~non-bold, bold lower case and bold upper case letters, respectively. The~set~of real numbers is denoted by $\setR$ and the complex plane is shown by $\setC$. $\mH^{\her}$, $\mH^{*}$ and $\mH^{\trp}$ indicate the Hermitian,~complex~conjugate and transpose of $\mH$, respectively. $\log\left(\cdot\right)$ is the binary logarithm. We denote the statistical expectation by $\E$, and the non-negative part of $x$ by $[x]^+=\max\{0, x\}$. The beta distribution with~the~shape parameters $\alpha$ and~$\beta$~is denoted by $\distB{\alpha}{\beta}$; moreover, $\man(\eta, \sigma^2)$ and $\mathcal{CN} (\eta, \sigma^2)$ represent the real and complex~Gaussian~distribution with mean $\eta$ and variance $\sigma^2$, respectively.
\section{Problem Formulation}
\label{sec:sys}
We consider downlink transmission in a multiuser \ac{mimo} wiretap setting. In this setting, a \ac{bs} with $M$ antennas intends to transmit confidential messages to $K$ legitimate users while the channel being overheard by an eavesdropper. For simplicity, we assume that the receiving terminals, i.e., the legitimate users and the eavesdropper, are single-antenna. By a same approach taken in \cite{asaad2017tas}, the analysis can be extended to scenarios with multi-antenna receiving terminals. The \ac{bs} is equipped with $L\leq M$ \ac{rf}-chains. Hence, in each coherence time interval, only $L$ transmit antennas are set active.

The uplink channel from user terminal $k$ to the \ac{bs}~is~represented by $\bh_k \in\setC^M$ and reads
\begin{align}
\bh_k=\sqrt{\beta_k} \hspace*{.7mm} \bg_k
\end{align}
where $k\in\set{1, \ldots,K}$ denotes the legitimate users and $k=\ee$ indicates the eavesdropper. The entries of $\bg_k \in\setC^M$ denote fast fading coefficients between user $k$ and the transmit antennas, and ${\beta_k}$ models path loss and shadowing. It is assumed that $\beta_k$ is constant over several coherence time intervals and is known in priori. This is the case in most practical scenarios. As the result, the legitimate uplink channel can be written as
\begin{align}
\mH=[\bh_1, \ldots, \bh_K]=\mG \mGam^{1/2}
\end{align}
where $\mG=[\bg_1, \ldots, \bg_K]$ and $\mGam$ is a $K\times K$ diagonal matrix with $[\mGam]_{kk}=\beta_k$. We further assume that the system operates in standard \ac{tdd} mode meaning that the uplink and downlink channels are reciprocal.

At the beginning of each coherence time interval, the~\ac{bs} employs a selection algorithm to select $L$ transmit antennas. We denote the set of the selected antennas with 
\begin{align}
\setL=\set{\ell_1, \ldots,\ell_L}
\end{align}
where $1\leq \ell_j \leq M$ for $j=1,\ldots,L$. The effective legitimate uplink channel after antenna selection is therefore described by $\tmH\in \setC^{L\times K}$ which is constructed from $\mH$ by collecting the $\ell_j$-th rows of $\mH$ for $j=1, \ldots, L$. Similarly, the effective uplink channel from the eavesdropper to the \ac{bs} is denoted by $\tbh_\ee\in \setC^L$ whose entries are the entries of $\bh_\ee$ indexed by $\setL$.
%to be active using a given . Denoting the channel matrix between the \ac{bs} and the legitimate receivers with $\mH$

\subsection{Secure Transmission under Antenna Selection}
For $k=1, \ldots, K$, let $u_k$ represent the confidential message aimed to be received by legitimate user $k$. $u_k$ is encoded into the codeword $[s_k(1), \ldots, s_k(N)]$ where $N$ is the code-length. The encoded vector $\bs(n)=[s_1(n), \ldots, s_K(n)]^\trp$ is then given to the \ac{bs} for transmission in the $n$-th time instant over the $L$ selected transmit antennas. For this aim, the \ac{bs} constructs the transmit signal $\bx(n)\in\setC^L$ from $\bs(n)$ using a linear precoder. This means that, for $n\in\set{1,\ldots,N}$,
%
%Denoting the encoded message of the legitimate user $k$ with $s_k$, the \ac{bs} intends to transmit the vector $\bs=[s_1, \ldots, s_K]^\trp$ confidentially over the wiretap channel.
%
%
%For this goal, the messages are individual
%
%
%
%%is equipped with an $M$ transmit antennas and all the receiving terminals are single-antenna. \ac{bs}
%\subsection{System Model}
%Let $\mH \in \setC^{M \times K}$ represent the uplink channel between the legitimate user terminals and the \ac{bs}. Moreover, denote the channel between the eavesdropper and the transmit antenna array with $\bg \in \setC^{M \times 1}$. The \ac{bs} selects the $L$ transmit antennas indexed by $\setJ=\{j_1, \cdots, j_{L}\}$ to be active. These antennas are selected as $\setJ = \mas(\mH, L)$ where $\mas$ is the \ac{tas} protocol. The transmit vector $\bx\in \setC^{L\times1}$ is constructed from the encoded messages $\bs$ via a linear precoder which means that
\begin{align}
\label{Eq:3}
\bx(n)=\sqrt{P}\hspace{0.7mm} \mW \bs(n),
\end{align}
where $P$ constrains the~transmit~power and $\mW \in \setC^{L \times K}$ is the shaping matrix satisfying $\E{\tr{\mW \mW^\her}}=1$. Assuming that $s_k(n)\sim \mathcal{CN}(0,1)$ for $k\in\set{1,\ldots,K}$, the total transmit power in \eqref{Eq:3} is constrained to $P$. %The transmit vector $\bx$ is then transmitted on the active antennas. %Denoting the signal at the legitimate terminal $k$ with $y_k$, 

By transmitting $\bx(n)$ over the selected antennas, the legitimate terminals receive
\begin{align}
\label{Eq:1}
\by(n)=\tmH^\trp \bx(n)+\bn_\mathrm{m}(n),
\end{align}
where $\by(n)\coloneqq[y_1(n), \ldots, y_K(n)]^\trp$ with $y_k(n)$ denoting the signal received by the $k$-th legitimate user in the time instant $n$, and $\bn_\mathrm{m}(n) \in \setC^{K \times 1}$ is \ac{iid} zero-mean complex Gaussian noise with variance $\sigma_\mm^2$. %$\tmH\in \setC^{L\times K}$ represents the effective channel matrix corresponding to the active transmit antennas whose row vectors are the rows of $\mH$ being indexed by $\setJ$. 
The eavesdropper moreover receives
\begin{align}
\label{Eq:2}
z(n)=\tbh_\ee^\trp \bx(n)+ n_\mathrm{e} (n)
\end{align}
where $n_\ee (n) \sim \mathcal{CN}(0,\sigma_\ee^2)$. %and $\tbg\in \setC^{L\times 1}$ denotes the effective eavesdropper channel corresponding to the selected antennas in $\setJ$, i.e., $\tbg = [\rmg_{j_1}, \ldots, \rmg_{j_L}]^\trp$. %The entry $\rmg_m$ denotes the $m$th entry of the eavesdropper channel $\bg$, which is modeled as a zero-mean and unit-variance \ac{iid} complex Gaussian random variable. We assume that the eavesdropper is passively overhearing the legitimate terminals which means that the \ac{csi} is not available at the eavesdropper. The \ac{csi} is however considered to be available at the legitimate terminals.

\subsection{Achievable Secrecy Rate}
\label{sec:Ach_rate}
In the absence of the eavesdropper, the maximum achievable rate of user $k$ is bounded from below by \cite{caire2010multiuser}
\begin{align}
\label{Eq:4}
\mar^\mm_k=\log (1+\sinr^\mm_k)
\end{align}
where $\sinr^\mm_k$ is the \ac{sinr} at the legitimate terminal $k$ and is given by
\begin{align}
\label{Eq:4-2}
\sinr^\mm_k = \frac{ \rho_\mm \abs{\tbh^\trp_k \bww_k}^2  }{ 1 + \rho_\mm \sum\limits_{j=1, j\neq k}^K \abs{\tbh^\trp_k \bww_j}^2 }.
\end{align}
Here, $\rho_\mm=P/\sigma_\mm^2$, and $\tbh_k$ and  $\bww_k$ are $L\times 1$ vectors which denote the $k$th column of $\tmH$ and $\mW$, respectively.

For user $k$, the maximum achievable secrecy rate is defined as the maximum rate at which the \ac{bs} can transmit information to legitimate user $k$ such that no information about $u_k$ is leaked to the eavesdropper. This rate is lower-bounded by~\cite{oggier2011secrecy}
\begin{align}
\label{Eq:4-1}
\mar^\rms_k=\left[ \mar^\mm_k - \mar^\ee_k \right]^+  
\end{align}
where $\mar^\ee_k$ represents the maximum achievable rate over the eavesdropper's channel under the worst-case assumption that the eavesdropper is able to cancel out the interference of other legitimate terminals\footnote{This is not necessarily the case, and therefore, \eqref{Eq:4-1} gives a lower bound.}. Consequently, $\mar^\ee_k$ is given by
\begin{align}
\label{Eq:4-SS}
\mar^\ee_k=\log (1+\snr^\ee_k)
\end{align}
where $\snr^\ee_k$ is the \ac{sinr} at the eavesdropper while overhearing $u_k$ and reads
\begin{align}
\snr^\ee_k =  \rho_\ee \abs{\tbh_\ee^\trp \bww_k}^2 
\end{align}
with $\rho_\ee = P/\sigma_\ee^2$.
%Here, $\snr^\ee_k$ is the \ac{sinr} at the eavesdropper while overhearing the $k$th user message considering the worst-case scenario in which the eavesdropper can successively cancel the interference of other legitimate terminals. More generally, one can define $\sinr^\ee_k (\alpha)$ for some $0\leq \alpha \leq 1$ to be the effective \ac{sinr} at the malicious terminal when the eavesdropper overhears the message of user $k$ and cancels the fraction $\alpha$ of the interference caused by other legitimate users. In this case,
%\begin{align}
%\sinr^\ee_k (\alpha) = \dfrac{ \rho_\ee \abs{\tbg^\trp \bww_k}^2  }{ 1 + (1-\alpha) \rho_\ee \sum\limits_{j=1, j\neq k}^K \abs{\tbg^\trp \bww_j}^2 }
%\end{align}
%where $\rho_\ee=P/\sigma_\ee^2$. Consequently, $\snr_k^\ee= \sinr^\ee_k (1)$ which reduces to
%\begin{align}
%\snr^\ee_k =  \rho_\ee \abs{\tbg^\trp \bww_k}^2 .
%\end{align}
By substituting into \eqref{Eq:4-1},  $\mar^\rms_k$ reads
\begin{align}
\label{Eq:last--RS}
\mar^\rms_k=\log \left( \frac{ 1+ \sinr_k^\mm }{ 1+ \snr_k^\ee }  \right).
\end{align}
\subsection{Relative Secrecy Cost}
Secrecy at the physical layer is obtained at the expense of reduction in the data rate. As \eqref{Eq:4-1} indicates, this cost depends on the quality of the eavesdropper's channel. We quantify this cost by defining the measure ``relative secrecy cost'' as follows. 
\begin{definition}[Relative secrecy cost]
\label{def:rel_cost}
%Consider a multiuser \ac{mimo} wiretap setting with $M$ transmit and $L$ active antennas. 
Assume $L$ antennas are selected out of $M$ available transmit antennas. Let $\mar^\rms_k(M,L)$ and $\mar^\mm_k(M,L)$ denote the secrecy rate to user $k$ and the achievable rate to user $k$ in the absence of the eavesdropper, respectively. The relative secrecy cost for user $k$ is~defined~as
\begin{align}
\sfC_k (M,L) \coloneqq 1-\dfrac{ \mar^\rms_k (M,L) }{\mar^\mm_k (M,L)}.
\end{align}
\end{definition}
From the definition of the relative secrecy cost, one simply observes that $0 \leq \sfC_k (M,L) \leq 1$ where the lower bound holds when $\mar^\mm_k (M,L) = \mar^\rms_k (M,L)$ and the upper bound is achieved when $\mar^\rms_k (M,L)=0$. $\sfC_k (M,L)$ determines the fraction of available rate being used to secure the transmission. We show that this cost converges to zero~in~massive \ac{mimo} wiretap setups even under antenna selection. We refer to this phenomenon as ``secrecy for free'' in massive \ac{mimo} systems in the presence of passive eavesdroppers.
%To avoid high complexity due to the required matrix inversion for \ac{zf} and \ac{mmse} precoding especially for large dimension \cite{gao2016energy}, in this paper,  
%This property is considered to be an asymptotic property which holds in the large-system limit.
\section{Secrecy For Free}
Secrecy for free intuitively means that the achievable rate in the absence of the eavesdropper and the achievable secrecy rate in the presence of the eavesdropper are nearly the same when the number of transmit antennas grows large. To formulate this property, we state the following definition. %this property, we make use of the former definition of relative secrecy cost. 
\begin{definition}[Asymptotic secrecy for free]
\label{def:sec_free}
Let $\rho_\mm = P/\sigma_\mm^2$ and $\rho_\ee = P/\sigma_\ee^2$ be bounded from above. For a given number of active transmit antennas $L$, the multiuser \ac{mimo} wiretap setting with passive eavesdropper illustrated in Section \ref{sec:sys} is said to asymptotically achieve secrecy for free when
\begin{align}
\lim_{M\uparrow \infty} \sfC_k (M,L) = 0
\end{align}
for any $k \in \set{1,\ldots,K}$.
\end{definition}
In a multiuser \ac{mimo} system, in which secrecy is achieved asymptotically for free, $\mar^\rms_k \approx \mar^\mm_k$ when the number of transmit antennas is large. This means that the \ac{bs} can confidentially transmit messages to each legitimate user with almost no loss in terms of the achievable rate.

In what follows, we show that this property holds in general in massive \ac{mimo} wiretap setups with passive eavesdroppers even when only a fixed number of transmit antennas are set active. Throughout our investigations, we consider a Rayleigh fast fading model for the channels. This means that the entries of $\bg_k$ for $k\in\set{1, \ldots, K,\ee}$ are independent complex Gaussian with zero mean and unit variance. We moreover consider \ac{mrt} precoding at the \ac{bs}, which is typical for massive \ac{mimo} systems \cite{marzetta2010noncooperative}. This means that we set $\bww_k={\tbh^*_{k}}/{\Vert \tbh_{k}\Vert}$ for $k\in\set{1, \ldots, K}$. The analysis is readily extended to other linear precoders as well as bi-unitarily invariant channel matrices\footnote{The random matrix $\mH\in\setC^{M\times K}$ is \textit{bi-unitarily invariant}, if for any pair of independent unitary matrices $\mU\in\setC^{M\times M}$ and $\mV\in\setC^{K\times K}$, the entries of $\mH$ and $\mU\mH\mV^{\her}$ have same distribution\cite{tulino2004random}.} by a similar~approach.
\subsection{Secrecy For Free under Full Transmit Complexity}
\label{sec:full}
We begin  with the case of full transmit complexity in which all transmit antennas are set active. In this case, $\tbh_k=\bh_k$ for $k\in\set{1, \ldots, K,\ee}$ and $\bww_k={\bh_{k}^*}/{\Vert \bh_{k}\Vert}$. Thus, the \ac{sinr} at the legitimate terminal $k$ reads
\begin{subequations}
\begin{align}
\hspace*{-3mm}\frac{1}{M}\sinr^\mm_k  &= \frac{1}{M} \dfrac{ \rho_\mm \norm{\bh_k}^2  }{ 1 +  \rho_\mm \sum\limits_{j=1, j\neq k}^K \dfrac{\abs{\bh_k^\trp \bh^*_j}^2}{\norm{\bh_j}^2} }\\
&= \frac{1}{M}\dfrac{\dfrac{\rho_\mm}{M} \beta_k \norm{\bg_k}^2  }{ \dfrac{1}{M} +  \rho_\mm \sum\limits_{j=1, j\neq k}^K \dfrac{\frac{1}{M^2} {\beta_k }\abs{\bg_k^\trp \bg^*_j}^2}{\frac{1}{M} \norm{\bg_j}^2} } \\
&\stackrel{\dagger}{\longrightarrow} \rho_\mm \beta_k
\label{eq:sinr_full}
\end{align}
\end{subequations}
where $\longrightarrow$ indicates convergence in mean square, and $\dagger$ comes from channel hardening \cite{hochwald2004multiple} and the favorable propagation property \cite{ngo2014aspects} of massive \ac{mimo} systems. In fact, from channel hardening, we have $\Vert\bg_k \Vert^2 / M \longrightarrow \E\{ \Vert\bg_k \Vert^2 \}/M=1$ as $M$ grows large. Moreover, the favorable propagation property of the channel for large $M$ implies that
$\abs{\bg_k^\trp \bg_j^*} / M \longrightarrow 0$ for any $k\neq j$. 

Denoting the achievable rate with $M$ transmit antennas over the legitimate channel $k$ by\footnote{We have dropped the argument $L$ for the case of full transmit complexity as in this case $L=M$.} $\mar^\mm_k(M)$, we conclude from \eqref{eq:sinr_full} that
\begin{align}
\mar^\mm_k(M) - \log \left( 1+ \rho_\mm \beta_k M \right) \longrightarrow 0. \label{eq:Rm_full}
\end{align}

Similarly, by invoking channel hardening and the favorable propagation property, we have for $\snr^\ee_k$
\begin{subequations}
\begin{align}
\frac{1}{M} \snr^\ee_k &= \frac{1}{M} \rho_\ee \dfrac{\abs{\bh_\ee^\trp \bh_k^*}^2}{\norm{\bh_k}^2} \\
&=  \dfrac{\dfrac{\rho_\ee}{M^2} {\beta_\ee } \abs{\bg_\ee^\trp \bg_k^*}^2}{\dfrac{1}{M} \norm{\bg_k}^2} \longrightarrow 0.
\label{eq:snr_e_full}
\end{align}
\end{subequations}
From \eqref{eq:sinr_full} and \eqref{eq:snr_e_full}, we can write %implies that
\begin{align}
\label{eq:NESBAT}
\left( \frac{ 1+ \sinr_k^\mm }{ 1+ \snr_k^\ee }  \right) / \left( 1+ \rho_\mm \beta_k M \right) \longrightarrow 1.
\end{align}
Considering \eqref{Eq:last--RS}, \eqref{eq:NESBAT} leads us to conclude that
\begin{align}
\mar^\rms_k(M) - \log \left( 1+ \rho_\mm \beta_k M \right) \longrightarrow 0. \label{eq:Rs_full}
\end{align}
Here, $\mar^\rms_k(M)$ is the achievable secrecy rate with full transmit complexity. From \eqref{eq:Rm_full} and \eqref{eq:Rs_full}, we have %in Definition \ref{def:rel_cost}, we have
\begin{align}
\lim_{M\uparrow\infty} \sfC_k(M) &=1 - \lim_{M\uparrow\infty} \frac{\log \left( 1+ \rho_\mm \beta_k M \right)}{\log \left( 1+ \rho_\mm \beta_k M \right)}= 0 \label{eq:cost_full}
%1 - \lim_{M\uparrow\infty} \frac{\log \left( 1+ \rho_\mm \beta_k M \right)}{\log \left( 1+ \rho_\mm \beta_k M \right)}
\end{align}
where $\sfC_k(M)$ represents the relative secrecy cost under full complexity, i.e., $L=M$. From \eqref{eq:cost_full}, one observes that secrecy in massive \ac{mimo} wiretap settings with passive eavesdroppers is achieved for free under full complexity, and therefore, for large $M$, we have $\mar_k^\rms \approx \mar_k^\mm$. The intuition behind this result is that in this setting, the precoder can accurately focus the transmission beam on the legitimate terminals, due to the large number of transmit antennas. This beam becomes significantly narrow, as the number of transmit antennas grows large, and therefore, the leakage to the eavesdropper vanishes. %due to the narrow beam of the transmitter.

\begin{remark}
Here, secrecy for free is achieved by simple \ac{mrt} precoding. This means that in a massive \ac{mimo} setting, the \ac{bs} can exclude the eavesdropper without knowing the channel state information of the eavesdropper.
\end{remark}

\subsection{Secrecy For Free under Antenna Selection}
\label{sec:tas}
With a large number of transmit antennas, two scenarios for antenna selection can be considered:%transmit antennas can be selected in two different cases:
\begin{enumerate}
\item The number of selected antennas $L$ grows large with the total number of antennas $M$, such that $L/M$ is kept fixed.
\item The number of selected antennas is kept fixed, e.g. $L=1$, while the total number of antennas growing large.% $M$ is growing. 
\end{enumerate}
In the first scenario, by a similar approach as in Section~\ref{sec:full}, it is shown that secrecy is asymptotically achieved for free. In fact, in this case, as the number of active antennas grows proportional with the total number of transmit antennas, the precoder can narrow its beam toward the legitimate users, and thus, the leakage to the eavesdropper vanishes in the large-system limit\footnote{To show this argument, one can start with a random selection algorithm which selects a fixed fraction of transmit antennas at random. Taking exactly same steps as in Section~\ref{sec:full}, it is  shown that secrecy is also achieved for free in this case. As the result, other algorithms achieve secrecy asymptotically for free in the first scenario, since all algorithms are superior to random selection.}. We therefore concentrate on the second scenario in this section and show that even for a fixed number of active antennas, secrecy is achieved asymptotically for free. %in \ac{mimo} wiretap settings.

For the sake of presentation, we set $K=1$. Nevertheless, all results and findings extend to arbitrary $K$ in a straightforward way. In this case, $\mH = \bh_1 \sqrt{\beta_1} \bg_1$ where we use the notation
\begin{subequations}
\begin{align}
\bh_1= [h_1 , \ldots, h_M]^\trp, \\
\bg_1= [g_1 , \ldots, g_M]^\trp.
\end{align}
\end{subequations}
To select a subset of transmit antennas, the following algorithm is employed at the beginning of each coherence~time~interval: The \ac{bs} sorts channel coefficients  $h_1,\ldots, h_M$ such that
\begin{align}
\abs{h_{\ell_1}}^2 \geq \ldots \geq \abs{h_{\ell_M}}^2
\end{align}
and selects the $L$ strongest antennas, i.e., $\setL=\set{\ell_1, \ldots, \ell_L}$.

Under this antenna selection algorithm, the \ac{sinr} at the legitimate terminal reads
%\begin{subequations}
\begin{align}
\sinr_1^\mm  &= {\rho_\mm} {  \norm{\tbh}^2  } = {\rho_\mm} \beta_1 { \sum_{\ell=1}^L \abs{g_{j_\ell}}^2  } = {\rho_\mm} \beta_1 \Xi
\end{align}
%\end{subequations}
where we define
\begin{align}
\Xi \coloneqq { \sum_{\ell=1}^L \abs{g_{j_\ell}}^2  }.
\end{align}
In the context of order statistics \cite{arnold1992first}, $\Xi$ is known as a trimmed sum. The large-system distribution of a trimmed sum has been given in \cite{stigler1973asymptotic}. Note that by the large-system limit in this case, we mean that $L$ is kept fixed and only $M$ grows large. In \cite{stigler1973asymptotic}, the asymptotic distribution of a trimmed sum is derived for a general distribution of summands. Noting that the summands in $\Xi$ are exponentially distributed, one can invoke the main theorem of \cite{stigler1973asymptotic} and write
\begin{align}
\Xi \sim \man \left( L\left( 1 + \psi \log \frac{M}{L} \right) , L \left( 2-\frac{L}{M} \right) \right)  \label{eq:Xi}
\end{align}
where $\psi = 1 / \log e \approx 0.6931$. From \eqref{eq:Xi}, one observes that as $M$ grows large, $\Xi$ converges in distribution to a Gaussian random variable whose mean increases with $\log M$ and whose variance converges to a constant. Consequently, one can write
\begin{subequations}
\begin{align}
&\lim_{M\uparrow \infty} \E \set{\frac{\sinr_1^\mm }{\log M}} =  \psi \rho_\mm \beta_1 L \\
&\lim_{M\uparrow \infty} \E \set{ \left\vert \frac{ \sinr_1^\mm }{\log M} - \E \set{\frac{ \sinr_1^\mm }{\log M}} \right\vert^2} =  0
\end{align}
\end{subequations}
which implies that 
\begin{align}
\frac{\sinr_1^\mm }{\log M} \longrightarrow  \psi \rho_\mm \beta_1 L \label{eq:SINR_m_asy}
\end{align}
or equivalently
\begin{align}
\mar_1^\mm (M,L) - \log \left( 1+ \psi \rho_\mm \beta_1 L \log M \right) \longrightarrow 0 \label{eq:Rm_sel}
\end{align}
where $\mar_1^\mm (M,L)$ denotes the achievable rate over the main channel when $L$ out of $M$ available transmit antennas are selected via the selection algorithm.

Considering the eavesdropper, $\snr_1^\ee$ is written as
\begin{subequations}
\begin{align}
\snr_1^\ee  &= {\rho_\ee} {  \frac{\abs{\tbh_\ee^\trp \tbh_1^*}^2}{\norm{\tbh_1}^2}  } \\
&= {\rho_\ee}\beta_\ee \norm{\tbg_\ee}^2 \cos^2 \theta
\end{align}
\end{subequations}
where $\theta$ denotes the Hermitian angle between $\tbg_1$ and $\tbg_\ee$ and is defined as
\begin{align}
\theta \coloneqq \cos^{-1} \left( \frac{\abs{\tbg_\ee^\trp \tbg_1^*}}{ \norm{\tbg_\ee} \norm{\tbg_1} } \right).
\end{align}
To determine the large-system limit of $\snr_1^\ee$, we consider the following lines of justifications:
\begin{itemize}
\item[(a)] As $\bh_1$ and $\bh_\ee$ are statistically independent, any ordered sorting on the entries of $\bh_1$ results in a random permutation of $\bh_\ee$. Therefore, the entries of $\tbh_\ee$ are statistically similar to the entries of a random selection. In other words, from the eavesdropper's point of view, the antenna selection algorithm is random selection. This fact implies that the entries of $\tbg_\ee$ are independent complex Gaussian with zero mean and unit variance.
\item[(b)] Note that the illustrated selection algorithm sorts the channel coefficients with respect to the magnitudes. This fact, implies that the direction of $\tbg_1$ is statistically similar to the direction of a randomly selected vector. Therefore, the Hermitian angle between $\tbg_1$ and $\tbg_\ee$ is distributed similar to the hermitian angle between two independent Gaussian random vectors\footnote{In the large-system limit, this statement can be rigorously justified~by~considering a sequence $\set{\theta_M}$ and showing that it converges in distribution to the Hermitian angle of two independent Gaussian vectors of the same size as $M$ grows large. Similar discussions can be found in \cite{asaad2017tas}.} of size $L$.
\item[(c)] The distribution of the squared cosine of the Hermitian angle between two independent Gaussian vectors of length $L$ is $\distB{1}{L-1}$; see \cite[Appendix C]{muller2001multiuser}. 
\item[(d)] Since $\tbg_\ee\sim\mathcal{CN}(\boldsymbol{0}, \mI_L)$, the normalization $\bv_\ee=\tbg_\ee/\norm{\tbg_\ee}$ is independent of $\norm{\tbg_\ee}$. Therefore, the random variable $\cos \theta = \abs{\bv_\ee^\trp \tbg_1^*}/\norm{\tbg_1}$ is independent of $\norm{\tbg_\ee}$, as well.
\end{itemize}
Considering the above lines of justifications, one can conclude that $\cos^2 \theta \sim \distB{1}{L-1}$ and is independent~of~$\norm{\tbg_\ee}^2$. $\norm{\tbg_\ee}^2$ is moreover a chi-square random variable with $2L$ degrees of freedom, mean $L$ and variance $L$. Consequently, the expected value of $\snr_1^\ee$ reads
\begin{subequations}
\begin{align}
\E \set{\snr_1^\ee }&= {\rho_\ee} \beta_\ee \hspace*{.9mm} \E \set{\norm{\tbg_\ee}^2} \E\set{\cos^2 \theta} \\
&= {\rho_\ee} \beta_\ee \ L \ \frac{1}{L} = {\rho_\ee} \beta_\ee.\label{eq:SNRe_mean}
\end{align}
\end{subequations}
To determine the variance of $\snr^\ee$, we note that
\begin{subequations}
\begin{align}
\E \set{\abs{\snr_1^\ee}^2} &= {\rho_\ee^2}\beta_\ee^2 \hspace*{.9mm} \E \set{\norm{\tbg}^2}^2 \E\set{\cos^2 \theta}^2 \\ %= 2{\rho_\ee^2} \beta_\ee^2
&= {\rho_\ee^2}\beta_\ee^2 \ L(L+1) \ \frac{2}{L(L+1)} = 2{\rho_\ee^2} \beta_\ee^2
\end{align}
\end{subequations}
which results in 
\begin{align}
\E \set{\abs{\snr_1^\ee}^2}- \E \set{\snr_1^\ee}^2 &= {\rho_\ee^2}\beta_\ee^2. \label{eq:SNRe_var}
\end{align}
From \eqref{eq:SNRe_mean} and \eqref{eq:SNRe_var}, it is concluded that $\snr_1^\ee$ takes random values around $\rho_\ee\beta_\ee$ with the finite variance $\rho^2_\ee\beta_\ee^2$. Thus, %Consequently, 
\begin{subequations}
\begin{align}
&\lim_{M\uparrow \infty} \E \set{\frac{\snr_1^\ee }{\log M}} =  0, \label{eq:SNR_e_avg}\\
&\lim_{M\uparrow \infty} \E \set{ \left\vert \frac{ \snr_1^\ee}{\log M} - \E \set{\frac{\snr_1^\ee}{\log M}} \right\vert^2}  =  0, \label{eq:SNR_e_var}
\end{align}
\end{subequations}
or equivalently, we can write
\begin{align}
\frac{\snr_1^\ee }{\log M} \longrightarrow 0. \label{eq:SNR_asymp}
\end{align}
Considering \eqref{eq:SINR_m_asy} along with \eqref{eq:SNR_asymp}, we have
\begin{align}
\mar_1^\rms (M,L) - \log \left( 1+ \psi \rho_\mm \beta_1 L \log M \right) \longrightarrow 0 \label{eq:Rs_sel}
\end{align}
where $\mar_1^\rms (M,L)$ denotes the achievable secrecy rate when $L$ transmit antennas are selected.

Invoking the arguments in \eqref{eq:Rm_sel} and \eqref{eq:Rs_sel}, the asymptotic relative secrecy cost $\sfC_1(M,L)$ in this case reads
\begin{align}
\hspace*{-1mm}\lim_{M\uparrow\infty} \sfC_1(M,L) \hspace*{-.5mm} = \hspace*{-.5mm} 1 \hspace*{-.5mm} - \hspace*{-.5mm} \lim_{M\uparrow\infty} \frac{\log \left( 1+ \psi \rho_\mm L \log M \right)}{\log \left( 1+ \psi \rho_\mm L \log M \right)} =0. \label{eq:SecFree_sel}
\end{align}
From \eqref{eq:SecFree_sel}, it is observed that even by selecting a \textit{fixed} number of antennas, secrecy is achieved for free as the \textit{total} number of available antennas grows unboundedly large. This observation intuitively comes from this fact that the growth in the total number of transmit antennas improves the channel~quality of the selected antennas while the eavesdropper's channel remaining unchanged. Hence, as $M$ grows large, the leakage to the eavesdropper becomes negligible compared to the achievable rate over the main channel, and we have $\mar^\rms \approx \mar^\mm$.
%In the previous section, we have shown that when the number of antennas at the transmit side increases, the secrecy achievable rate tends to the achievable rate to the main channel. 

\begin{remark}
The large-system analysis considers a fixed~number of active antennas. This means that the result is valid~even when a \textit{single} antenna is selected. The characterization for single transmit antenna selection can be more precisely addressed using the extreme value distribution given by Fisher-Tippet law \cite{arnold1992first}; see the studies in \cite{bai2009rate,asaad2017asymptotic} for~some~particular~examples. Nevertheless, both the approaches lead to this conclusion that the secrecy cost vanishes as $M$ grows large.
\end{remark}

\begin{remark}
\label{remark:speed}
Considering the cases of full transmit complexity and antenna selection with fixed $L$, one observes that in either cases $\mar^\ee_k$ does not grow large with $M$ while the achievable rate over the main channel scales in terms of $M$. The growth in the achievable rate $\mar^\mm_k$ is however of different orders in these cases: With full complexity, the growth order is\footnote{Note that $\sinr_k^\mm$ grows linearly with $M$ in this case; see \eqref{eq:sinr_full}.} $\log M$, while under antenna selection, $\mar^\mm_k$ grows proportional to\footnote{Under antenna selection, $\sinr_k^\mm$ grows with $\log M$; see \eqref{eq:SINR_m_asy}.} $\log\log M$. This fact indicates that the secrecy cost converges to zero with lower speed, when a fixed number of antennas are selected.
\end{remark}

\begin{remark}
\label{rmrk:2}
By either changing the precoding scheme, e.g. zero forcing scheme, or using a superior selection algorithm, e.g. algorithm in \cite{bereyhi2018iterative}, the secrecy performance in this setting is improved. This fact indicates that although the analyses are given for \ac{mrt} precoding and a specific selection algorithm, the results guarantee the achievability of secrecy for free for a large class of algorithms as well as other precodings.
\end{remark}

\section{Numerical Investigations}
\begin{figure}[t]
\centering
\includegraphics[scale=1]{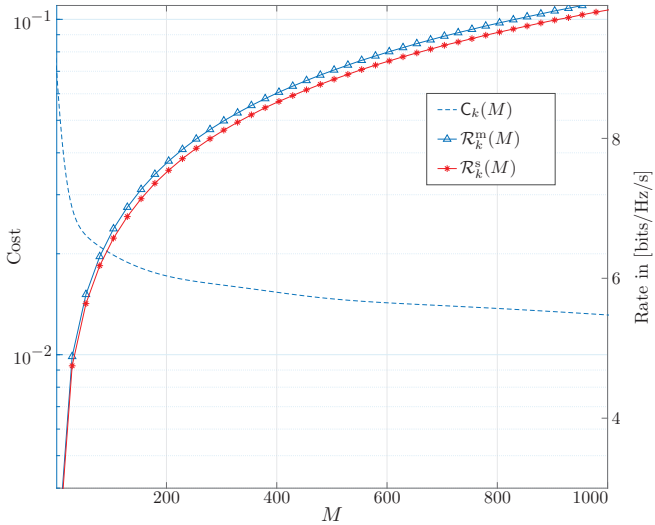}
%\vspace*{1mm}
%\hspace*{-.8cm}  
%\resizebox{1.15\linewidth}{!}{
%\pstool[width=.35\linewidth]{Fig_final/Fig1-neu.eps}{
%\psfrag{rate}[c][b][0.23]{$\rm Rate \ in \ [bits/Hz/s]$}
%\psfrag{M}[c][c][0.25]{$M$}
%\psfrag{Cost}[c][t][0.23]{$\rm Cost$}
%
%\psfrag{COSTFULLAA}[l][c][0.22]{$\hspace*{-8mm}\sfC_k(M)$}
%\psfrag{RMMMFULLAI}[l][c][0.22]{$\hspace*{-8mm}\mar^\mm_k(M)$}
%\psfrag{RSSSFULLAA}[l][c][0.22]{$\hspace*{-8mm}\mar^\rms_k(M)$}
%%%y-axis
%\psfrag{0}[r][c][0.22]{$0$}
%\psfrag{0.05}[r][c][0.22]{$0.05$}
%\psfrag{0.1}[r][c][0.22]{$0.1$}
%\psfrag{0.15}[r][c][0.22]{$0.15$}
%\psfrag{0.2}[r][c][0.22]{$0.2$}
%\psfrag{0.25}[r][c][0.22]{$0.25$}
%\psfrag{0.3}[r][c][0.22]{$0.3$}
%\psfrag{0.35}[r][c][0.22]{$0.35$}
%\psfrag{0.4}[r][c][0.22]{$0.4$}
%\psfrag{0.45}[r][c][0.22]{$0.45$}
%\psfrag{0.5}[r][c][0.22]{$0.5$}
%
%%x-axis
%\psfrag{1}[l][c][0.22]{$1$}
%\psfrag{2}[l][c][0.22]{$2$}
%\psfrag{3}[l][c][0.22]{$3$}
%\psfrag{4}[l][c][0.22]{$4$}
%\psfrag{5}[l][c][0.22]{$5$}
%\psfrag{6}[l][c][0.22]{$6$}
%\psfrag{7}[l][c][0.22]{$7$}
%\psfrag{8}[l][c][0.22]{$8$}
%\psfrag{9}[l][c][0.22]{$9$}
%%
%\psfrag{0}[c][b][0.22]{$1$}
%\psfrag{100}[c][b][0.22]{ }
%\psfrag{200}[c][b][0.22]{$200$}
%\psfrag{300}[c][b][0.22]{ }
%\psfrag{400}[c][b][0.22]{$400$}
%\psfrag{500}[c][b][0.22]{ }
%\psfrag{600}[c][b][0.22]{$600$}
%\psfrag{700}[c][b][0.22]{ }
%\psfrag{800}[c][b][0.22]{$800$}
%\psfrag{900}[c][b][0.22]{ }
%\psfrag{1000}[r][b][0.22]{$1000$}
%\psfrag{-1}[r][t][0.2]{$-1$}
%\psfrag{-2}[r][t][0.2]{$-2$}
%\psfrag{10}[r][b][0.22]{$10\hspace*{3mm}$}
%\psfrag{x}[l][b][0.22]{$10$}
%}}
\caption{The relative secrecy cost, achievable rate and secrecy rate for each legitimate user versus the number of antennas at the \ac{bs}. Here, $K=4$, $\log \rho_\mm=0$ dB and $\log \rho_\ee=-10$ dB.\vspace*{3.2mm}} %As the figure shows, $\mar^\rms_k$ closely tracks $\mar^\mm_k$ and the relative secrecy cost $\sfC_k(M)$ vanishes as $M$ grows large.\vspace*{-1mm}}
\label{fig:1}
\end{figure}

To confirm our analyses, we consider some~numerical~examples. Fig.~\ref{fig:1} illustrates the robustness against passive eavesdropping under full transmit complexity, i.e., $L=M$. Here, $K=4$ legitimate users and a single eavesdropper is considered. The channels are \ac{iid} Rayleigh fading with zero mean and unit variance. It is assumed that the users are uniformly distributed in the cell, and the path loss is compensated at the receiving terminals, i.e., $\beta_k=1$ for all $k$. The \ac{snr} at each legitimate terminal and the eavesdropper has been set to $\log \rho_\mm=0$~dB and $\log \rho_\ee=-10$ dB, respectively.  For this setting, the achievable rate over the legitimate channel $\mar^\mm_k(M)$ as well as the achievable secrecy rate for~each~legitimate user $\mar^\rms_k(M)$ has been plotted as a function of the number of antennas at the \ac{bs}\footnote{Note that in this case, due to the uniform distribution of the users in the cell, $\mar^\mm_k=\mar^\mm_j$ and $\mar^\rms_k=\mar^\rms_j$ for all $k\neq j$.} $M$.

As Fig.~\ref{fig:1} shows, the achievable secrecy rate closely tracks the rate achieved over the legitimate channel in the absence of the eavesdropper, i.e., $\mar^\mm_k(M)$. The relative secrecy cost $\sfC_k(M)$ has been further sketched in Fig.~\ref{fig:1}. One should note that $\sfC_k(M)$ is \textit{relative},~meaning that it does not directly scale with $\mar^\mm_k(M)-\mar^\rms_k(M)$, but with the ratio  $[\mar^\mm_k(M)-\mar^\rms_k(M) ]/\mar^\mm_k(M)$. As it is observed in the figure, the relative secrecy cost drops rapidly with respect to the number transmit antennas which confirms the analysis in Section~\ref{sec:full}.

%the achievable rate over the legitimate channel $\mar^\mm$ along with the achievable secrecy rate $\mar^\rms$ have been plotted for the both cases of full complexity and \ac{tas}.

\begin{figure}[t]
\centering
\includegraphics[scale=1]{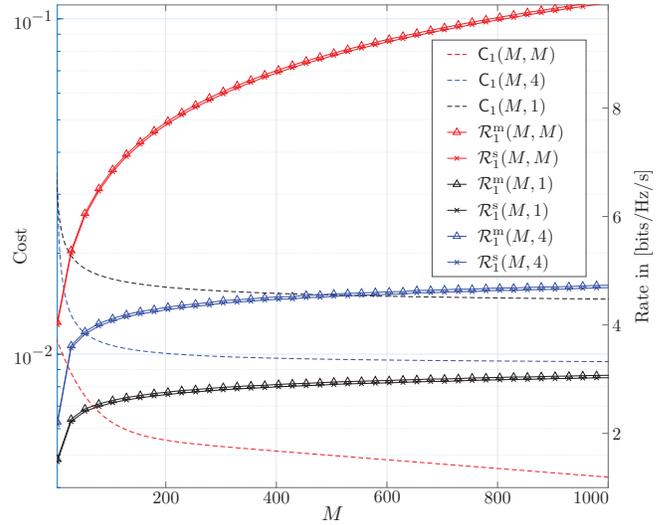}
\caption{The relative secrecy cost, achievable rate and secrecy rate~versus the total~number of transmit antennas $M$ for different number of \ac{rf}-chains, i.e., $L=1$ and $L=4$, at the \ac{bs}. Here, $K=1$, $\log \rho_\mm=0$ dB and $\log \rho_\ee=-15$ dB.} %As the figure shows, in the both cases $\mar^\rms$ closely tracks $\mar^\mm$ as $M$ grows large. The relative secrecy cost under full complexity $\sfC(M)$ vanishes faster than $\sfC(M.L)$ indicating the case with \ac{tas}.}
\label{fig:2}
\end{figure}

To investigate the robustness against passive eavesdropping under antenna selection, we have considered a \ac{mimo} wiretap setting with a single legitimate terminal and an eavesdropper in Fig.~\ref{fig:2}. Here, the same channel model as in Fig.~\ref{fig:1} is assumed. Moreover, the \ac{snr}s at the legitimate terminal and the eavesdropper have been set $\log \rho_\mm=0$~dB and $\log \rho_\ee=-15$~dB, respectively. In this figure, the achievable rates $\mar_1^\mm(M,L)$ and $\mar_1^\rms(M,L)$, as well as the relative secrecy cost $\sfC_1 (M,L)$, have been plotted against the total number of transmit antennas\footnote{Note that the curves for the case of antenna selection start on the $x$-axis from $M=L$, since we have $M\geq L$.}~$M$ for two cases of single transmit antenna selection, i.e. $L=1$, and $L=4$. It is assumed that the algorithm illustrated in Section~\ref{sec:tas} is employed for antenna selection. For the sake of comparison the results for full transmit complexity have been sketched as well.

As Fig.~2 depicts even with a single active antenna at~the \ac{bs}, the achievable rate over the legitimate channel and the secrecy rate meet at large values of $M$. From the figure, it is observed that under antenna selection, the relative secrecy cost converges to zero slower than the case with full transmit complexity. In fact, the slope of $\sfC_1(M,M)$ at large values of $M$ is considerably larger than the slope of $\sfC_1(M,1)$ and $\sfC_1(M,4)$. As discussed in Remark~\ref{remark:speed}, this observation comes from the two following facts:
\begin{inparaenum}
\item The achievable rate $\mar_k^\mm(M,L)$, for a fixed $L$, grows large significantly slower than $\mar_k^\mm(M,M)$, i.e., the rate achieved by full transmit complexity\footnote{In fact, $\mar_k^\mm(M,L)$, for a fixed $L$, grows proportional to $\log\log M$, while $\mar_k^\mm(M,M)$ grows with $\log M$.}, and 
\item the information leakage under antenna selection does not vanishes as fast as the case with all transmit antennas being active.
\end{inparaenum}

%\begin{figure}[t]
%\centering
%\input{Fig_final/figure2.tex}
%\caption{The achievable rate and the secrecy rate versus the total~number of transmit antennas $M$ for different number of \ac{rf}-chains, i.e., $L=1$ and $L=4$, at the \ac{bs}. Here, $K=1$, $\log \rho_\mm=0$ dB and $\log \rho_\ee=-15$ dB.}
%\end{figure}

\section{Conclusion}
This paper has studied the robustness of massive \ac{mimo} wiretap settings against passive eavesdropping. Our investigations have shown that even when the \ac{bs} employs a \textit{fixed} number of its transmit antennas, including the case with a \textit{single} active antenna, the information leakage to the eavesdropper vanishes as the \textit{total} number of transmit antennas grows large. This fact indicates that in massive \ac{mimo} systems, regardless of the number of active antennas, secrecy is achieved almost ``for free''. Our analytic results guarantee the robustness of massive \ac{mimo} settings against passive eavesdropping for a large class of selection algorithms and precodings.

From numerical simulations, it is known that in contrast~to setups with \textit{passive} eavesdroppers, massive \ac{mimo} systems are not robust against \textit{active} eavesdropping \cite{kapetanovic2015physical}. The large-system characterization of \ac{mimo} wiretap settings under active eavesdropping attacks is therefore an interesting direction for future work. The work in that direction is~currently~ongoing.

\bibliography{ref}
\bibliographystyle{IEEEtran}

\begin{acronym}
\acro{mimo}[MIMO]{Multiple-Input Multiple-Output}
\acro{csi}[CSI]{Channel State Information}
\acro{awgn}[AWGN]{Additive White Gaussian Noise}
\acro{iid}[i.i.d.]{independent and identically distributed}
\acro{ut}[UT]{User Terminal}
\acro{bs}[BS]{Base Station}
\acro{mt}[MT]{Mobile Terminal}
\acro{eve}[Eve]{Eavesdropper}
\acro{tas}[TAS]{Transmit Antenna Selection}
\acro{lse}[LSE]{Least Squared Error}
\acro{rhs}[r.h.s.]{right hand side}
\acro{lhs}[l.h.s.]{left hand side}
\acro{wrt}[w.r.t.]{with respect to}
\acro{tdd}[TDD]{Time-Division Duplexing}
\acro{rsb}[RSB]{Replica Symmetry Breaking}
\acro{papr}[PAPR]{Peak-to-Average Power Ratio}
\acro{mrt}[MRT]{Maximum Ratio Transmission}
\acro{zf}[ZF]{Zero Forcing}
\acro{rzf}[RZF]{Regularized Zero Forcing}
\acro{snr}[SNR]{Signal to Noise Ratio}
\acro{sinr}[SINR]{Signal to Interference plus Noise Ratio}
\acro{rf}[RF]{Radio Frequency}
\acro{mf}[MF]{Match Filtering}
\acro{mmse}[MMSE]{Minimum Mean Squared Error}
\end{acronym}

\end{document}